\documentclass[twocolumn,showpacs,preprintnumbers,amsmath,amssymb,prb,aps]{revtex4-1}
\usepackage{graphicx}
\usepackage{bm}
\usepackage{amsmath}
\usepackage{multirow}

\begin{document}
\title{Structural properties of Fe-Ni/Cu/Fe-Ni trilayers on Si~(100)}
\author{Ananya Sahoo , Maheswari Mohanta, S. K. Parida}
\author{V. R. R. Medicherla}
\email{Corresponding author: mvramarao1@gmail.com}
\affiliation{Department of Physics, ITER, Siksha 'O' Anusandhan Deemed to be University, Bhubaneswar 751030, India}
\date{\today}
\begin{abstract}
We investigate the structural properties of Fe$_{1-x}$Ni$_x$/Cu/Fe$_{1-x}$Ni$_x$ ( $x=0.5$, non Invar and  $x=0.36$, Invar) trilayers deposited on Si~(100)~at room temperature using dc magnetron sputtering technique in ultra high vacuum conditions taking high purity Fe, Ni and Cu metals with Cu layer thickness  4, 6 and 8 nm for each alloy composition. The structure of the alloy films of the trilayers was investigated using x-ray diffraction and the thickness \& roughness of the layers were obtained by x-ray reflectivity measurement. Both the as prepared and annealed trilayers exhibit layered structure. The as deposited Fe-Ni alloy in non Invar trilayer exhibits only fcc structure whereas in Invar alloy it exhibits a mixed fcc and bcc phases. Interestingly after annealing at 425$^0$C in ultra high vacuum, the Invar alloy completely transformed to  fcc structure for all Cu thicknesses. In both Invar and non Invar trilayers, the Bragg reflections corresponding to Fe-Ni alloy layers become sharp after annealing. The induced structural transformation in Invar trilayer is explained using enhanced diffusion of Fe and Ni atoms at high temperatures. 
\end{abstract}

\maketitle
\section{Introduction}
Ferromagnetic (FM) multilayers are considered as important materials by physicists and engineers in the recent past due to the novel and exciting physical properties such as giant magnetoresistance (GMR)\cite{baibich, binach, fullerton, locatelli}, tunneling magnetoresistance (TMR) \cite{ikeda}, magnetic anisotropy \cite{rizal}, surface plasmon resonance and giant magneto-reflectivity (GMRE)\cite{Lodewijks, maksymov}, they exhibit. The fundamental physics of the novel magnetoresistive phenomena in magnetic multilayers is still elusive\cite{camley, levy}. The magnetic multilayers can be used in panoply of technological applications. The best application of them is in the fabrication of devices that utilize the spin of the electron\cite{slonc}. One such device called spin valve can be used as read head for computer hard disks. The spin valves can also be used in non-volatile random access memory structures. The emergence of magnetoelectronics can bring a sea change in electronics industry. The magnetic multilayers play a key role in improving the data storage density. A recent review by Rizal {\textit et.al} \cite{conrad-review} describes in detail the fabrication and characterization of the magentic multilayers consisting of alternate stacks of FM and nonmagnetic (NM) layers. The observed GMR in magnetic multilayers was attributed to spin dependent scattering at the interface\cite{baibich,camley}. The spin dependent scattering at the bulk of the FM layer was also included to obtain GMR along with interracial scatting\cite{barnas, sauren}.\\

Pure metal magnetic multilayers such as Co/Cu are immiscible, whereas the Fe-Ni/Cu multilayers are not immiscible but show interdiffusion\cite{hecker}. The magnetic multilayers based on Fe-Ni layers were at the centre of focus for researchers primarily due to the Invar property exhibited by Fe-Ni 36\% alloy. On the other hand the high Ni concentration Fe-Ni alloys exhibit very high permeability and are called permalloys which are used in magnetic shielding. Fe-Ni thin films are soft magnetic materials with high anisotropic magnetoresistance (AMR), high saturation magnetization and low coercivity\cite{myung, andri, kuru, robert}. The crystal structure and magnetism of Fe-Ni thin films strongly depend on Ni content\cite{kojima, ustinov}. The physical properties of Fe-Ni films are also known to vary with the  surface morphology and texture of the films\cite{malkin, cao}. The onset of interdiffusion in thick FeNi and Cu multilayers starts at around 200$^0$C~\cite{bruckner}. Many researchers have focused their attention on permalloy, Fe$_{0.2}$Ni$_{0.8}$ multilayers \cite{hecker, elefant, bruckner}, but the work reported on multilayers of low Ni concentration fcc alloys are scarce\cite{elefant, kim}.\\

The physical properties of magnetic multilayers made of Fe-Ni alloys pose challenges both in basic physics and technological points of view. In particular Fe-Ni/Cu/Fe-Ni trilayers with Fe-Ni in the Invar composition is more complicated as Invar Fe-Ni alloy does not expand or contract thermally, but Cu layer does. Such a differential thermal expansion at the interface strongly depends on thickness and roughness of various layers of the stack.The contraction/expansion at the interface can lead to new crystallographic phase formation. The method of deposition and temperature of annealing play a key role in deciding the crystal structure and surface - interface characteristics of multilayers. The structural parameters such as layer roughness is important for device functioning.\\
 
In this paper we report the crystal structure of Fe-Ni layers of Fe-Ni/Cu/Fe-Ni trilayer system deposited on Si~(100)~using magnetron sputtering method.  The crystal structure of Fe-Ni alloys of the trilayers was studied using x-ray diffraction (XRD) technique. The composition of the alloys was studied using energy dispersive x-ray analysis (EDAX) and the thickness \& roughness of the layers were investigated by x-ray reflectivity. The results suggest that the gentle annealing at 425$^0$C improves the crystallinity of Fe-Ni layers and stabilizes fcc structure in all trilayers prepared. \\

\section{Experiment}
Trilayered thin films Fe$_{0.5}$Ni$_{0.5}$/Cu/Fe$_{0.5}$Ni$_{0.5}$ and Fe$_{0.64}$Ni$_{0.36}$/Cu/Fe$_{0.64}$Ni$_{0.36}$ were synthesized at room temperature using a dc magnetron sputtering system (Orion-8, AJA Int. Inc). Two distinct series of samples were prepared in each case with Cu layer thickness 4, 6 and 8~nm. Fe$_{0.64}$Ni$_{0.36}$ alloy lies in the Invar region of Fe-Ni alloys and Fe$_{0.5}$Ni$_{0.5}$ lies in the non Invar region. The thickness of Fe-Ni alloy was kept constant at 50~nm for all the trilayers prepared. Deposition was done on clean Si (100) substrate with the base pressure of 10$^{-8}$Torr. The substrate Si was rotated at 60~rpm to maintain better in plane homogeneity of the deposited layers. High purity (99.95\%)~Fe, Ni and Cu metal targets were used for deposition. The Target-Substrate distance was maintained at 40~mm during deposition. The rate of deposition and deposition time were calibrated using x-ray reflectivity spectra prior to each trilayer deposition. \\

All the trilayers were subjected to annealing at 425$^0$C using electron beam heating for 1~hr duration in a vaccum of 10$^{-7}$ Torr. The crystal structure of  as deposited and annealed films were obtained using x-ray diffraction (XRD) technique using CuK$_\alpha$ radiation. The elemental composition of as deposited Fe-Ni layers was confirmed by energy dispersive x-ray analysis (EDAX) using scanning electron Microscope(SEM). The layer thickness and surface/interface roughness were studied by X-ray reflectivity (XRR) technique. \\

\section{Results and Discussion}

\section{Results}
\begin{figure}[h!]
 \vspace{0ex}
\begin{center}
\includegraphics [scale=0.4]{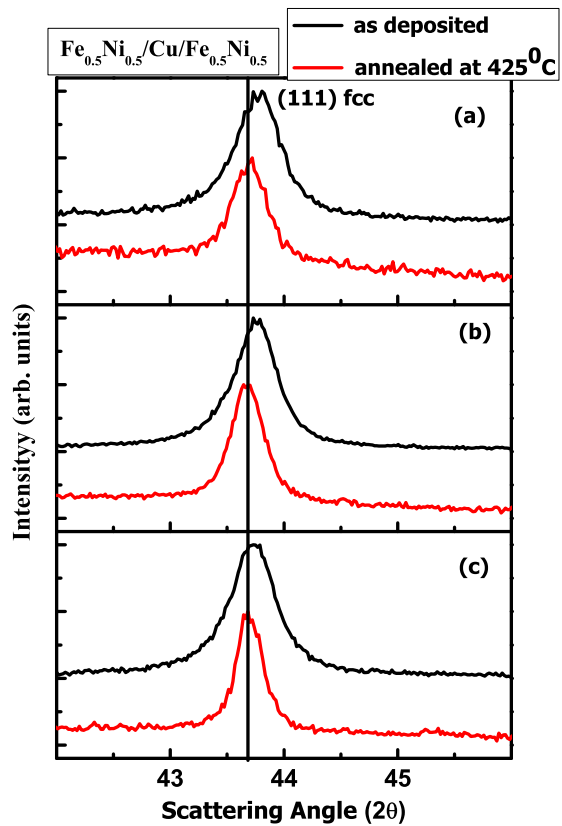}
 \vspace{-2ex}
\caption{XRD data of Fe$_{0.5}$Ni$_{0.5}$/Cu/Fe$_{0.5}$Ni$_{0.5}$ trilayers with Cu layer thickness (a) 4 nm (b) 6 nm (c) 8 nm. Only fcc~(111) peak is observed both in as deposited~(black line) and in annealed trilayers at 425$^0$C (red line).} 
\end{center}
 \vspace{0ex}
\end{figure}
The XRD data of as deposited and annlealed Fe$_{0.5}$Ni$_{0.5}$/Cu/Fe$_{0.5}$Ni$_{0.5}$ trilayers is shown figure 1 (a), (b) and (c) respectively for Cu layer thickness of 4, 6 and 8~nm. The most intense fcc~(111) peak of Fe$_{0.5}$Ni$_{0.5}$ is shown in each case. The fcc~(111) peak for as deposited trilayer occurs at around a scattering angle of 43.6$^0$. The scattering angle exhibits only a marginal variation of 0.2$^0$ in as deposited trilayers as shown in the figure. After annealing, fcc~(111) peak occurs at 43.6$^0$ and becomes narrow compared to that of as deposited. Peak narrowing upon annealing suggests the improved crystallinity of the trilayers. The crystallite size increased drastically after annealing as shown in table 1. \\

\begin{figure}[h!]
 \vspace{0ex}
\begin{center} 
\includegraphics [scale=0.4]{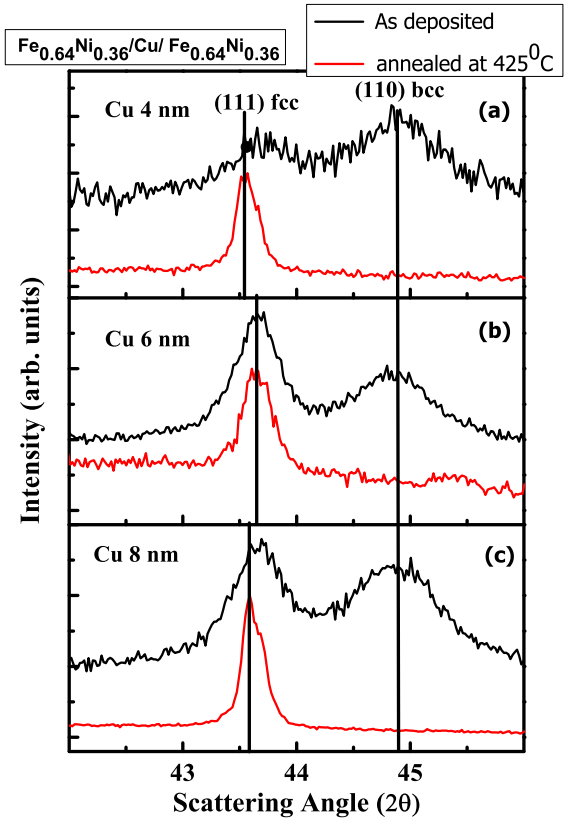}
 \vspace{-2ex}
\caption{XRD data of Fe$_{0.64}$Ni$_{0.36}$/Cu/Fe$_{0.64}$Ni$_{0.36}$ trilayers with Cu layer thickness (a) 4 nm (b) 6 nm (c) 8 nm. In as deposited trilayers~(black line) both fcc~(111) and bcc~(110) reflections are observed whereas after annealing at 425$^0$C, only fcc~(111) reflection is observed.}
\end{center}
 \vspace{0ex}
\end{figure}
Figure 2 (a), (b) and (c) show the XRD data of as deposited and annealed Fe$_{0.64}$Ni$_{0.36}$/Cu/Fe$_{0.64}$Ni$_{0.36}$ trilayers. Interestingly, these trilayers exhibit two Bragg reflections fcc~(111) and bcc~(110) respectively at 43.8$^0$ and 44.9$^0$ of scattering angle. The bcc~(110) reflection is much broader compared to that of fcc~(111) reflection. After annealing bcc~(110) reflection disappeared in all three trilayers as shown in figure 2. fcc~(111) peak becomes very sharp After annealing and occurs at 43.7$^0$ scattering angle.\\
\begin{table*}
\begin{center}
\caption{Lattice parameter, Crystallite size and Composition of of the alloy layers. Lattice parameter and crystallite size were obtained from XRD data. The composition of the alloy layers was obtained from EDAX measurement}
\begin{tabular}{|c|c|c|c|c|c|c|c|c|c|}
\hline
 & & \multicolumn{4}{ c |}{\bf Lattice Parameter~(\AA)} &  \multicolumn{2}{ c|} {\bf Crystallite Size~(nm)}&\multicolumn{2}{ c |}{\bf EDAX Composition}\\ 
\cline{3-10}
{\bf Trilayer}& {\bf Cu layer} & \multicolumn{2}{ c |} {\bf FCC Phase} & \multicolumn{2}{ c |}{\bf BCC Phase} & & & &\\ 
\cline{3-6}
& {\bf thickness} & {\bf As }    & {\bf Annealed}    & {\bf As } & {\bf Annealed}& {\bf As } & {\bf Annealed} & {\bf Fe} & {\bf Ni}\\
& {\bf (nm)} & {\bf deposited} & {\bf @425$^0$C} & {\bf deposited} & {\bf @425$^0$C }& {\bf deposited} & {\bf @425$^0$C} & {\bf content\%} & {\bf content\%}\\
\hline
& 4 & 3.574 & 3.582 & -  & -& 0.27 & 0.390 & 52.20 & 47.80\\
\cline{2-10}
Fe$_{0.5}$Ni$_{0.5}$ & 6 & 3.578 & 3.584& - & -& 0.29 & 0.40 &49.00 &50.10\\
\cline{2-10}
 & 8 & 3.585 & 3.586 & -& -& 0.27 & 0.48&50.48&49.52\\
\hline
 & 4 & 3.594& 3.594 & 2.856 & -& 0.17 & 0.59 & 57.42&42.58\\
\cline{2-10}
Fe$_{0.64}$Ni$_{0.36}$ & 6 & 3.590& 3.590& 2.860 & -& 0.28 & 0.45 &57.69&42.31\\
\cline{2-10}
 & 8 &3.591& 3.590& 2.858& -& 0.23 & 0.61&60.37&39.60\\
\hline
\end{tabular}
\end{center}
\end{table*}
The lattice parameters, the crystallite size and the composition of all the trilayers is shown in table 1. The lattice parameters were obtained by analyzing XRD data using POWDERX software.  The crystallite size was calculated using Debye-Sherrer formula taking (111) reflection for fcc phase and (110) reflection for bcc phase. The composition of the layers was measured using EDAX technique. After annealing, the lattice parameter of Fe$_{0.5}$Ni$_{0.5}$ alloy is about 3.58~\AA, on the other hand the lattice parameter of  Fe$_{0.64}$Ni$_{0.36}$ alloy is about 3.59~\AA. The lattice parameter of bcc phase observed in as deposited Fe$_{0.64}$Ni$_{0.36}$ alloy is about 2.86~\AA. These lattice parameter values are in close agreement with that of corresponding poly crystalline bulk alloys \cite{shakti}. In case of Fe$_{0.5}$Ni$_{0.5}$/Cu/Fe$_{0.5}$Ni$_{0.5}$ trilayers, the composition obtained from EDAX closely agrees with the deposited composition. Whereas for Fe$_{0.64}$Ni$_{0.36}$/Cu/Fe$_{0.64}$Ni$_{0.36}$ trilayers EDAX composition differs from the deposited composition by about 10\% as shown in the table. In all the trilayers deposited, the crystallite size increased after annealing. The lowest crystallite size of 0.17~nm was observed for as deposited Fe$_{0.64}$Ni$_{0.36}$/Cu/Fe$_{0.64}$Ni$_{0.36}$ trilayers with Cu thickness of 4 nm (refer table 1). \\

\begin{figure}[h!]
 \vspace{0ex}
\begin{center}
\includegraphics [scale=0.4]{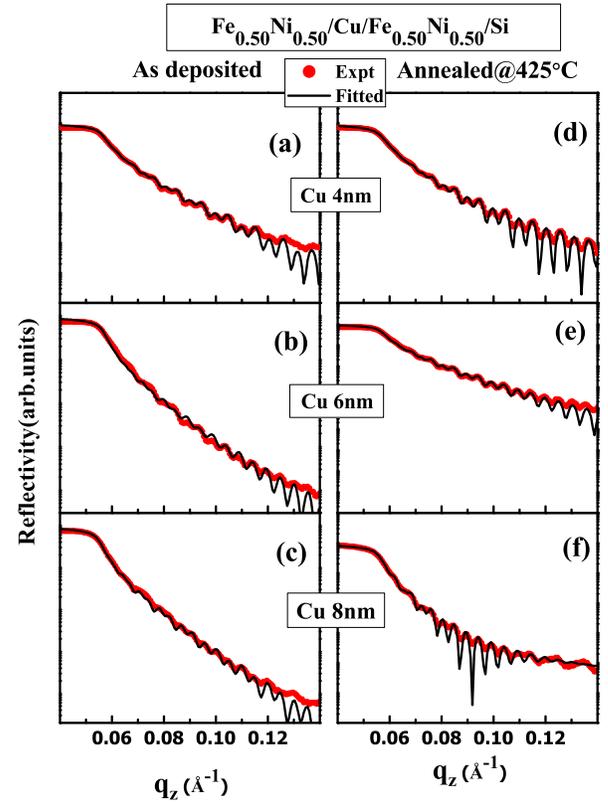}
 \vspace{-2ex}
\caption{XRR data of as deposited (left panels with Cu layer thickness (a) 4 nm, (b) 6 nm, (c) 8 nm) and annealed (right panels with Cu layer thickness (d) 4 nm (e) 6 nm (f) 8 nm ) Fe$_{0.5}$Ni$_{0.5}$/Cu/Fe$_{0.5}$Ni$_{0.5}$ trilayer. Symbol represents experimental data and line represent numerical fit using the Parratt formalism.}
\end{center}
 \vspace{0ex}
\end{figure}
X-ray reflectivity (XRR) patterns of both as deposited ((a), (b), (c)) and annealed ((d), (e), (f)) Fe$_{.5}$Ni$_{0.5}$/Cu/Fe$_{0.5}$Ni$_{0.5}$ trilayers are shown in Figure 2 ((a) to (f)). The experimental reflectivity data has been fitted using Parratt formalism\cite{parrot}. Kiessig fringes are clearly observed in both as deposited and annealed trilayers indicating a layered structure in both the cases. 
 
\begin{figure}[h!]
 \vspace{0ex}
\begin{center}
\includegraphics [scale=0.4]{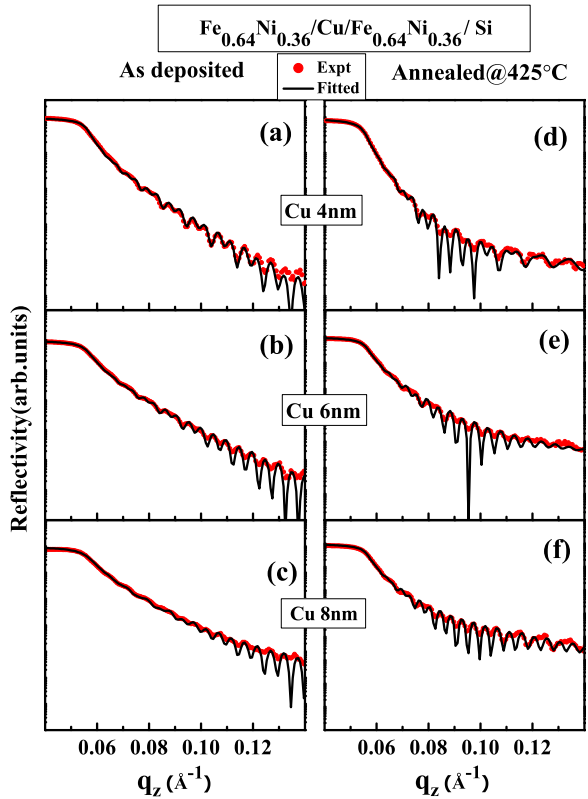}
 \vspace{-2ex}
\caption{XRR data of as deposited (left panels with Cu layer thickness (a) 4 nm, (b) 6 nm, (c) 8 nm) and annealed (right panels with Cu layer thickness (d) 4 nm (e) 6 nm (f) 8 nm ) Fe$_{0.64}$Ni$_{0.36}$/Cu/Fe$_{0.64}$Ni$_{0.36}$ trilayers. Symbol represents experimental data and line represents numerical fit using the Parratt formalism.}
\end{center}
 \vspace{0ex}
\end{figure}

Figure 3 ((a) to (f)) shows the x-ray reflectivity patterns of both as deposited ((a), (b), (c)) and annealed ((d), (e), (f)  Fe$_{0.64}$Ni$_{0.36}$/Cu/Fe$_{0.64}$Ni$_{0.36}$ trilayers. The experimental XRR data is fitted numerically using Parrot formalism in the same way as done for Fe$_{0.5}$Ni$_{0.5}$/Cu/Fe$_{0.5}$Ni$_{0.5}$ trilayers. In this trilayer also Kiessig fringes are observed indicating layered structure in both as deposited and annealed trilayers.  \\

Some general trends are observed in the XRR  patterns of both the trilayer systems prepared. For example, the x-ray intensity is found to decay faster in case of as deposited trilayers compared to that of annealed ones. In as deposited trilayers, the top Fe-Ni layer is expected to form clusters and thus a rough surface which strongly scatters x-ray and thus intensity decays faster. After annealing, the clusters diffuse and make a uniform layer to form a smooth surface giving rise to reduced x-ray scattering compared to as deposited ones. The XRR fitting parameters are shown in table 2 for both the trilayer systems. The thickness of each of the layers are in good agreement for all the trilayers as shown in the table.        
\begin{table*}
\begin{center}
\caption{Thickness and roughness of various layers of the as deposited \& annealed trilayers obtained by fitting x-ray reflectivity data using Parrat formalism}
\begin{tabular}{|c|c|c|c|c|c|c|}
\hline
{\bf Trilayer} & {\bf Cu layer} & {\bf Layer} & \multicolumn{2}{ c |}{\bf As deposited} & \multicolumn{2}{ c|} {\bf Annealed@425$^0$}\\
\cline{4-7}
& {\bf thickness} & {\bf of the stack} & {\bf Thickness} & {\bf Roughness} & {\bf Thickness} & {\bf roughness}\\
& {\bf (nm)} & & {\bf (nm)} & {\bf (nm)} & {\bf (nm)} & {\bf (nm)}\\
\hline
& & $Fe_{0.5}Ni_{0.5}$ & 49.42 & 1.7 & 50.6 & 1.7\\
\cline{3-7}
& 4 & Cu & 3.6 & 1.1 & 4.2 & 1.1\\
\cline{3-7}
& & $Fe_{0.5}Ni_{0.5}$ & 49.2& 2.5 & 48.5 & 1.2\\
\cline{2-7}
& & $Fe_{0.5}Ni_{0.5}$ & 51.5& 1.5& 49.5 & 1.6\\
\cline{3-7}
Non Invar & 6 & Cu & 5.6& 2 & 5.0 & 2\\
\cline{3-7}
& & $Fe_{0.5}Ni_{0.5}$ & 48.6& 2.5& 49.1 & 2\\
\cline{2-7}
& & $Fe_{0.5}Ni_{0.5}$ & 48.2 & 1.5& 47.2& 2.1\\
\cline{3-7}
& 8 & Cu & 7.0& 1.1 & 7.5 & 2.9\\
\cline{3-7}
& & $Fe_{0.5}Ni_{0.5}$ & 50.0 & 2.0 & 47.6& 2.8\\
\hline
& & $Fe_{0.64}Ni_{0.36}$ & 50.2 & 1.0 & 52.2 & 3.5\\
\cline{3-7}
& 4 & Cu & 3.6 & 2.0 & 4.0 & 2.2\\
\cline{3-7}
& & $Fe_{0.64}Ni_{0.36}$ & 51.0 & 1.0 & 52.2 & 0.7\\
\cline{2-7}
& & $Fe_{0.64}Ni_{0.36}$ & 50.2 & 1.6  & 48.9 & 2.6\\
\cline{3-7}
Invar & 6 & Cu & 5.9 & 1.4 & 6.0 & 2.5\\
\cline{3-7}
& & $Fe_{0.64}Ni_{0.36}$ & 51.1 & 1.9 & 48.4 & 2.4\\
\cline{2-7}
& & $Fe_{0.64}Ni_{0.36}$ & 49.8 & 1.0 & 50.9 & 2.5\\
\cline{3-7}
& 8 & Cu & 8.3 & 1.8 & 8.3 & 1.9\\
\cline{3-7}
& & $Fe_{0.64}Ni_{0.36}$ & 52.2 & 1.7 & 51.7 & 2.4\\
\hline
\end{tabular}
\end{center}
\end{table*}

The Fe$_{0.5}$Ni$_{0.5}$ alloy in the as deposited trilayer, Fe$_{0.5}$Ni$_{0.5}$/Cu/Fe$_{0.5}$Ni$_{0.5}$ forms in fcc structure and maintains it after annealing. In this trilayer system the crystallinity of Fe$_{0.5}$Ni$_{0.5}$ alloy layer improved after annealing as observed from the increase in crystallite size (ref. table 1). Whereas the Fe$_{0.64}$Ni$_{0.36}$ alloy in the as deposited trilayers Fe$_{0.64}$Ni$_{0.36}$/Cu/Fe$_{0.64}$Ni$_{0.36}$ exhibit both fcc and bcc phases and transforms completely to fcc phase after annealing. \\

Fe$_{1-x}$Ni$_x$ bulk alloys form in bcc structure upto a Ni concentration, $x\leq 0.26$ and transform to fcc structure for $x > 0.3$\cite{wasserman,shakti}. For x values ranging from 0.26 to 0.3, the alloys are in mixed phase containing both fcc and bcc phases. The mixed phase bulk alloy does not become single phase by thermal annealing\cite{shakti}. In case of thin films deposited on a substrate kept at room temperature, the scenario is quite different. The sputtered metals, Fe \& Ni are not distributed uniformly at microscopic  level on the substrate which leads to minute variations of composition of the alloy. The variation in composition gives rise to small variations in lattice parameter and thus the observed fcc~(111) and bcc~(110) reflections are broad in as deposited trilayers. After annealing composition become uniform and produce sharp fcc~(111) Bragg reflection for all trilayers. In non Invar, $x=0.5$ alloy, the small variations in composition will produce only fcc alloys, on the other hand the Invar, $x=0.36$ alloy exhibits both bcc and fcc phases as $x=0.36$ alloy is very close to the structural phase boundary. After annealing, the composition becomes uniform on the substrate and consequently the trilayer exhibits only fcc structure.\\

The diffusion of atomic species takes place when there is a spatial concentration gradient that is expected in as deposited trilayers. The particle flux density is related to the concentration gradient by Fick's law , ${\bf J}=-D{\bf \nabla} c$, where D is the diffusion coefficient and c, the concentration of the atomic species. The diffusion coefficient D strongly depends on temperature. If we assume Maxwell-Boltzman distribution for atoms, then the probability for an atom to attain an energy $E$ is given by $f(E)=Ae^{-E/k_BT}$, where A is a constant. If $E_d$ is the diffusion barrier energy then for $E>E_d$, predominant diffusion occurs.  In case of as deposited Invar trilayer, cluster formation with slight variation in composition is anticipated. Each cluster is of either bcc or fcc structure. At room temperature, the diffusion is negligible so that clusters remain stable maintaining the mixed fcc and bcc phases. When the trilayer is annealed, the atoms in the cluster vibrate with higher amplitude and have more energy to cross the diffusion barrier. This leads to strong diffusion across the clusters of different composition and structure. The diffusion continues as long as there is a concentration gradient of atomic species. In case Fe-Ni alloys, both Fe and Ni diffuse in opposite directions till the film attains a homogeneous composition. \\

\section{Conclusion}
In summary, we have prepared trilayers of Fe-Ni alloys with Cu as a spacer layer on clean Si~(100) substrate by dc magnetron sputtering method using high purity Fe, Ni and Cu metals. We have prepared Fe$_{0.5}$Ni$_{0.5}$/Cu/Fe$_{0.5}$Ni$_{0.5}$ non Invar trilayer and Fe$_{0.64}$Ni$_{0.36}$/Cu/Fe$_{0.64}$Ni$_{0.36}$ Invar trilayer by taking Cu layer thickness of 4 nm, 6 nm and 8 nm in each case. The trilayers have been investigated using XRD, XRR and EDAX techniques. The XRD results suggest that the as deposited non Invar trilayer contains Fe-Ni alloy in fcc phase whereas the as deposited Invar trilayer forms a mixed phase containing both  fcc and bcc phases of the Fe-Ni alloy. After annealing, bcc phase disappears in Invar trilayer and fcc~(111) peak becomes sharp in both Invar and non Invar trilayer. The transformation of bcc phase into fcc phase after annealing is explained using thermally activated diffusion process.

\section{Acknowledgement}
Authors acknowledge Dr.Mukul Gupta \& Dr. V. R. Reddy, UGC-DAE CSR, Indore for providing facilities for trilayer preparation \& characterization. AS would like to thank Mr. Layanta Behera, Mr. Anil Gome and Mr. A. K. Ahire respectively for trilayer deposition \& XRD, XRR and EDAX measurements. 


\clearpage
\begin{thebibliography}{99}
\bibitem{baibich}
M. N. Baibich, J. M. Broto, A. Fert, F. N. V. Dau, F. Petroff, P. Etienne, G. Creuzet, A. Friederich, J. Chazelas, Phys. Rev. Lett. {\bf 61}, 2472 (1988).
\bibitem{binach}
G. Binasch, P. Grünberg, F. Saurenbach, W. Zinn,  Phys. Rev. B {\bf 39}, 4828 (1989).
\bibitem{fullerton}
E. Fullerton, M. Conover, J. Mattson, C. Sowers, S. Bader,  Appl. Phys. Lett.  {\bf 63}, 1699 (1993).
\bibitem{locatelli}
N. Locatelli, V.  Naletov, J. Grollier, G. de Loubens, V. Cros, C. Deranlot, C. Ulysse, G. Faini,  Appl. Phys. Lett.,   {\bf 98}, 062501 (2011).
\bibitem{ikeda}
S. Ikeda, J. Hayakawa, Y. M. Lee, R. Sasaki, T. Meguro, F. Matsukura,H. Ohno,   Jpn. J. Appl. Phys. {\bf 44}, L1442 (2005).
\bibitem{rizal}
C. Rizal, B. Moa, J. Wingert, O. G. Shpyrko,  IEEE Trans. Magn. {\bf 2}, 2352932 (2015).
\bibitem{Lodewijks}
K. Lodewijks, N. Maccaferri, T. Pakizeh, R. K. Dumas, I. Zubritskaya, J.Akerman,  P.Vavassori, A. Dmitriev,  Nano Lett. {\bf 14} 7207 (2014).
\bibitem{maksymov}
I. S. Maksymov, Reviews Phys. {bf 1} 36 (2016).
\bibitem{camley}
R. E. Camley and J. Barnas, Phys. Rev. Lett. 63, 664 (1989).
\bibitem{levy}
P. M. Levy, K. Ounadjela, S. Zhang, Y. Wang, C. B. Sommers,
and A. Fert, J. Appl. Phys. 67, 5914 (1990).
\bibitem{slonc}
J. C. Slonczewski, Phys. Rev. 8 39, 6995 (1989).
\bibitem{conrad-review}
Conrad Rizal, Belaid Moa and Boris B. Niraula, magnetochemistry, 2016, {\bf 2}, 22.
\bibitem{barnas}
J. Barans, A. Fuss, R.E. Camely, P. Gr{\"u}nberg, W. Zinn, Phys. Rev. B {\bf 42}, 8110 (1990).
\bibitem{sauren}
F. Saurenbach, J, Barnas, G. Binasch, M. Vohl, P. Gr{\"u}nberg, W. Zinn, Thin Solid Films {\bf 175}, 317 (1989).
\bibitem{hecker}
M. Hecker, D. Tietjen, H. Wendrock, C. M. Schneider, N. Cramer, L. Malkinski, R. E. Camely, Z. Celinski, J. Magn. Magnetic Materials, {\bf 247}, 62 (2002).
\bibitem{myung}
N. V. Myung, D. Y. Park, B. Y. Yoo, P. T. A. Sumodjo, J. Magn. Magn. Mater. {\bf 265}, 189 (2003).
\bibitem{andri}
P. C. Andricacos and N. Robertson, IBM J. Res. Dev., , {\bf 42}, 671 (1998).
\bibitem{kuru}
H. Kuru, H. Kockar, M. Alper and O. Karaagac, J. Magn. Magn. Mater. {\bf 377}, 59 (2015).
\bibitem{robert}
N Robertson, B. H. L Hu and T. Ching, IEEE T. Magn. {\bf 33}, 2818 (1997).
\bibitem{kojima}
T. Kojima, M. Ogiwara, M. Mizuguchi, M. Kotsugi, T. Koganezawa, T. Ohtsuki, T. Tashiro and K. Takanashi, J. Phys.: Condens. Mat. {\bf 26}, 064207 (2014).
\bibitem{ustinov}
A. I. Ustinov, S. S. Polishchuk, S. A. Demchenkov and L. V. Petrushinets, J. Alloy.
Compd. {\bf 622}, 54(2015).
\bibitem{malkin}
L. M. Malkinski, R. Eskandari, A. L. Fogel and S. Min, J. Appl. Phys. {\bf 111},07A320 (2012).
\bibitem{cao}
Y. Z. Cao, Q. Wang, G. J. Li, J. J. Du, C. Wu and J. C. He, J. Magn. Magn. Mater. {\bf 332}, 38 (2013).
\bibitem{bruckner}
W. Br{\"u}ckner, S. baunack, M. Hecker, J-I. M{\"o}nch, L. V. Loyen, C. M. Schneider, Appl. Phys. Lett. {\bf 77}, 358 (2000).
\bibitem{elefant}
D. Elefant, R. Sch{\"a}fer, J. Thomas, H. Vinzelberg and C. M. Schneider, Phys. Rev B {\bf 77}, 014426 (2008).
\bibitem{kim}
J. Gi Kim, K. H. Han, S. Ho Song, A. Reilly, Thin Solid Films {\bf 440}, 54 (2003).
\bibitem{parrot}
L. G. Parratt,  Physical Review {\bf 95(2)}, 359 (1954).
\bibitem{wasserman}
E. F. Wassermann, in Ferromagnetic Materials, edited by H. J.Buschow and E. P. Wohlfarth,  Elsevier, Amsterdam,Vol. 5 (1990).
\bibitem{shakti}
S. S. Acharya, V. R. R. Medicherla, R. Rawat, K. Bapna, K. Ali, D. Biswas, and K. Maiti, J. Elect. Spect. Relat. Phenom. {\bf 212}, 1 (2016).
\end{thebibliography}
\end{document}